# Numerical investigation of the scale effects of rock bridges

Fengchang Bu[1,2,3], Lei Xue[1,2,3*], Mengyang Zhai[1,2,3], Chao Xu[1,2,3], Yuan Cui[1,2,3]

***Corresponding author.**

**E-mail address of all co-authors:**

bfc@mail.iggcas.ac.cn (Fengchang Bu)

xuelei@mail.iggcas.ac.cn (Lei Xue)

zhaimengyang@mail.iggcas.ac.cn (Mengyang Zhai)

xuchao@mail.iggcas.ac.cn (Chao Xu)

cuiyuan@mail.iggcas.ac.cn (Yuan Cui)

1. Key Laboratory of Shale Gas and Geoengineering, Institute of Geology and Geophysics, Chinese Academy of Sciences, Beijing 100029, China
2. Innovation Academy for Earth Science, Chinese Academy of Sciences, Beijing 100029, China
3. College of Earth and Planetary Sciences, University of Chinese Academy of Sciences, Beijing 100049, China

**Abstract**

The concept of joint persistence has been widely used to study the mechanics and failure processes of rock masses benefitting from the simplicity of statistical linear weighing of the discontinuity. Nevertheless, this term neglects the scale effects of rock bridges, meaning that the same joint persistence may refer to different numbers and spacings of rock bridges, leading to erroneous equivalent rock mass responses. To fill in this gap, an intact rock bridge was dispersed as multi rock bridges while maintaining a constant joint persistence, subjected to direct shear by conducting numerical simulations employing Universal Distinct Element Code (UDEC). In this way, scale effects of rock bridges were investigated from the perspective of load-displacement curves, stress and displacement fields, crack propagations and AE characterizations. Results revealed that the shear resistance and the area and value of stress-concentration decreased with increasing dispersion. Furthermore, uneven distribution of displacement fields in an arc manner moving and degrading away from the load was first observed, indicating the sequential failure of multi rock bridges. It was also found that the propagation of wing cracks was insensitive to scale, while the asperity of macro shear fracture mainly formed by secondary cracks decreased with increasing dispersion. In addition, increasing dispersion of rock bridges would overlap the failure precursors identified by intense AE activities. Based on the abovementioned results, we evaluated existing methods to characterize the joint persistence, and a threshold was observed to possibly define a rock bridge.

**Highlights**

- Mechanical properties of rock bridges deteriorated with decreasing relative scales.
- Scale effects of rock bridges were appropriately characterized using existing indexes.
- A possible threshold to define a rock bridge was found.

# 1 Introduction

Researchers have long recognized that the discontinuity consisting of joints and rock bridges within natural rock masses appears to have an important role in rock engineering (Terzaghi 1962; Lajtai 1969; Einstein et al. 1983; Stead and Eberhardt 2013). With higher shear strength than joints, rock bridges provide key resistance to reduce the potential for damage in rock masses to develop (Diederichs and Kaiser 1999; Cai et al. 2004; Hencher and Richards 2015; Jiang et al. 2015a). To evaluate the resistant capacity of rock bridges, Jennings (1970) introduced the concept of joint persistence, $K$, as a measure of rock mass discontinuity, defined as the ratio of the joint length (1D) or area (2D) of the joint surface and given by:

$$K = \frac{\sum J_i}{\sum J_i + \sum R_i} \quad (1)$$

where $J_i$ and $R_i$ were the ith length (area) of the joint and rock bridge, respectively (Fig. 1a). Accordingly, Jennings' criterion was proposed to compute the combined strength of joints and rock bridges:

$$\tau = K(c_j + \sigma \tan \varphi_j) + (1 - K)(c_r + \sigma \tan \varphi_r) \quad (2)$$

where $\tau$ and $\sigma$ was the shear stress and normal stress, $c_j$ and $c_r$ was the equivalent cohesion of joints and rock bridges, and $\varphi_j$ and $\varphi_r$ was the equivalent friction of joints and rock bridges, respectively.

Benefitting from the simplicity of statistical linear weighing of the discontinuity, Jennings' criterion had been widely used in the investigations of the mechanics and failure processes of rock bridges (Jamil 1992; Zhang et al. 2005; Alzo'ubi 2012; Bonilla-Sierra et al. 2015; Jiang et al. 2015b; Cylwik 2021; Zare et al. 2021). Nevertheless, Prudencio and Jan (2007) noticed that this criterion disregarded the impact of the joints on the stress field and assumed simultaneous failure of the rock bridges and the joints. Elmo et al. (2018) also pointed that the term, $K$, neglected block forming potential and scale effects, meaning that if the $K$-value was invariant while the numbers and spaces of rock bridges were variant, the strength of rock masses and failure processes would be the same according to Jennings' criterion, which did not square with the facts, as shown in Fig. 1. Consequently, the investigation of the scale effects of the rock bridges is desirable for rock engineering design considering the discontinuity.

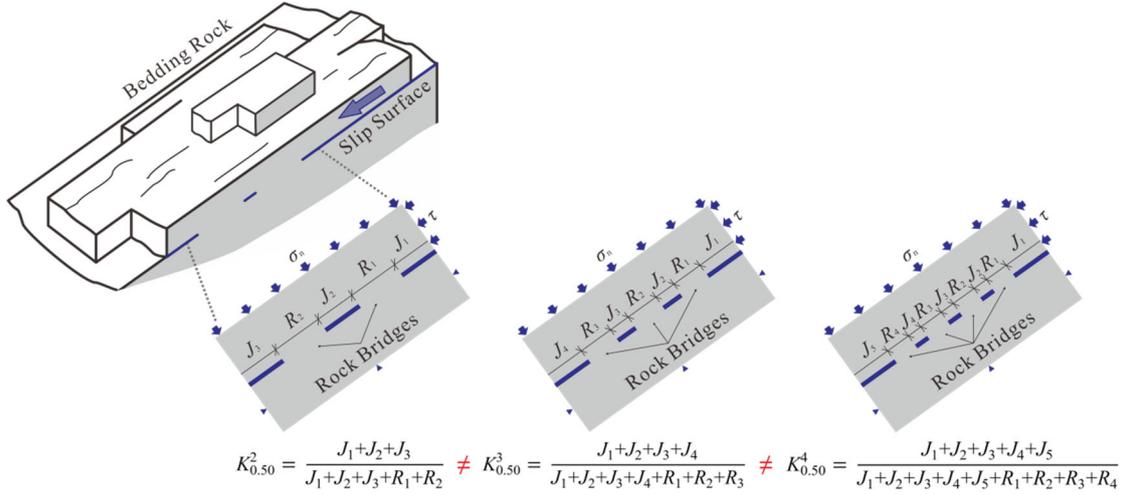

$$K_{0.50}^2 = \frac{J_1+J_2+J_3}{J_1+J_2+J_3+R_1+R_2} \neq K_{0.50}^3 = \frac{J_1+J_2+J_3+J_4}{J_1+J_2+J_3+J_4+R_1+R_2+R_3} \neq K_{0.50}^4 = \frac{J_1+J_2+J_3+J_4+J_5}{J_1+J_2+J_3+J_4+J_5+R_1+R_2+R_3+R_4}$$

**Fig. 1** Schematic view of rock bridges in a rock slope engineering and the calculation of joint persistence, $K_{\text{joint persistence}}^{\text{number of rock bridges}}$, bringing the limitation that the same joint persistence may refer to different scales of rock bridges, leading to erroneous equivalent rock mass responses

However, it appears to be difficult to investigate the scale effects of rock bridges in the field and laboratory due to the invisibility of rock bridges inside natural rock masses (Shang et al. 2017). Researchers always make a back analysis of pre-existing intact rock bridges to determine the resistance (Dershowitz and Einstein 1988), and a classical research method is the direct shear test with a constant normal load applied to the discontinuous plane because of the consistency of boundary conditions between laboratory tests and engineering practices (e.g., rock slope stability and surface excavation stability) (Lajtai 1969; Muralha et al. 2014). In this context, though a plethora of researchers investigated the effects of normal stress (Bagheripour and Mostyn 1995; Wong et al. 2001; Cundall et al. 2016; Yang and Kulatilake 2019), joint length (Zhang et al. 2005; Ghazvinian et al. 2007; Asadizadeh et al. 2018; Tang et al. 2020), joint orientation (Gehle and Kutter 2003; Gerolymatou and Triantafyllidis 2016; Hokmabadi et al. 2016; Zhong et al. 2020) and joint overlap (Savilahti 1991; Kemeny 2004; Sarfarazi et al. 2014) on rock bridges in direct shear, rare researchers investigate the scale effects because of the high cost and high failure rate of the handcraft rock bridge specimens (Shang et al. 2018). Thus, it is necessary to introduce numerical simulations into the investigation of scale effects of rock bridges.

The main objectives of this paper are to study the scale effects of rock bridges in direct shear by numerical simulation. Section "Method" showed the methodology, calibration and simulation in detail.

Section "Results" displayed the scale effects of rock bridges from the perspective of load-displacement curves, stress and displacement fields, crack propagations and AE characterizations. Previous approaches to describe the joint persistence were evaluated and discussed in the "Discussion". "Conclusions" were given finally.

## 2  Method

### 2.1  Discrete element modelling

The discrete element modelling (DEM) introduced by Cundall and Strack (1979) has been widely used in the analysis of rock mechanics problems because of its ability to explicitly represent fractures and bond failure of rocks. One of the mainstream commercial software based on the DEM is the Universal Distinct Element Code (UDEC) developed by the ITASCA Consulting Group (USA), in which a rock material is modelled as an assembly of blocks bonded with contacts, as shown in Fig. 2a. These blocks respond according to Newton's second law, and contacts among them are prescribed by a force-displacement law (Itasca, 2014), as shown in Fig. 2b. In other words, the microproperties of these blocks and contacts determine the macroscale mechanical behaviours of the numerical models. Normally, the bulk modulus ($K^{block}$) and shear modulus ($G^{block}$) of blocks and normal stiffness ($k_n$) and shear stiffness ($k_s$) of contacts determine the deformation behaviours, and cohesive ($c^{cont}$), friction angle ($\varphi^{cont}$) and tensile strength ($\sigma_t^{cont}$) of contacts determine the strength. In detail, in the direction perpendicular to a contact surface, the stress-displacement relation is assumed to be linear (Itasca, 2014):

$$\Delta\sigma_n = - k_n \Delta u_n \tag{3}$$

where $\Delta\sigma_n$ is the effective normal stress increment and $\Delta u_n$ is the normal displacement increment. If $\sigma_t^{cont}$ is exceeded by $\sigma_n$, the latter equals zero. In the other direction tangential to a contact surface, the stress-displacement relation is always assumed to be controlled by the Coulomb friction law:

$$|\tau_s| \leq c^{cont} + \sigma_n \tan\varphi^{cont} = \tau_{max} \tag{4}$$

where $\tau_s$ is the shear stress, then:

$$\Delta\tau_s = - k_s \Delta u_s^e \tag{5}$$

however, if $|\tau_s| \geq \tau_{max}$, then:

$$\tau_s = \text{sign}(\Delta u_s^e)\, \tau_{max} \tag{6}$$

where $\Delta u_s^e$ is the elastic component of the incremental shear displacement and $\Delta u_s$ is the total incremental shear displacement. In this way, the failure processes of rocks can be simulated in UDEC.

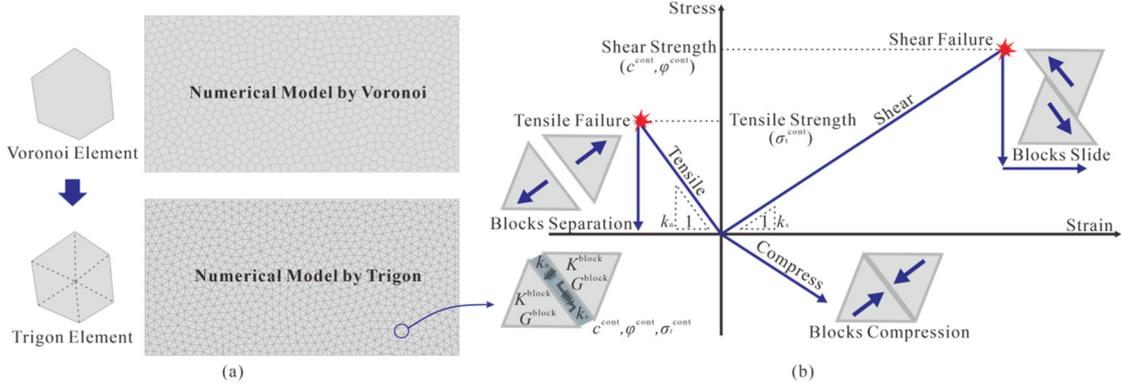

**Fig. 2 a** Illustration of the Voronoi block pattern modified to Trigon approach by defining triangular finite difference zones. **b** Constitutive behaviour in shear and tension

In the initial version of the UDEC, simulated cracks could only develop in a predetermined direction. To solve this problem, a polygon block pattern in UDEC was proposed by Lorig and Cundall (1989) and evolved into Voronoi later, allowing a more reasonable crack propagation in a numerical model. This Voronoi block pattern has been widely adopted in the simulations of rock behaviours (Christianson et al. 2006; Lan et al. 2010; Stavrou et al. 2019). Furthermore, Gao and Stead (2014) modified this traditional Voronoi block pattern to Trigon by dividing the polygonal blocks into triangular finite zones, as shown in Fig. 2a, allowing less dependence on the computational grid.

## 2.2 Physical prototype and calibration

Natural rock specimens are considered ideal for both experiments and simulations (Cao et al. 2019). In this paper, granitic rock specimens collected from a quarry in Qingdao City, Shandong Province of China were used as the physical prototypes (Yang and Kulatilake 2019). These granites were machined as cylindrical (diameter 50 mm × height 100 mm) and rectangular (length 200 mm × width 100 mm × height 100 mm) specimens to conduct uniaxial compression tests and direct shear tests, respectively. The former was used to estimate the macroscale mechanical properties like unconfined compression strength (UCS), and the latter was used to study the failure processes of rock bridges. Furthermore, as shown in Fig. 3, the rectangular specimens were processed into specimens with non-persistent open joints using a high-pressure water jet cutting machine, and the lengths of rock bridges were set as 125 mm ($K = 0.75$), 150 mm ($K = 0.50$) and 175 mm ($K = 0.25$). Then, the lower parts of the specimens were fixed, and a constant normal load and an invariant left velocity boundary were subjected to the upper parts. More details related to the experiments are available in the reports by Yang and Kulatilake (2019).

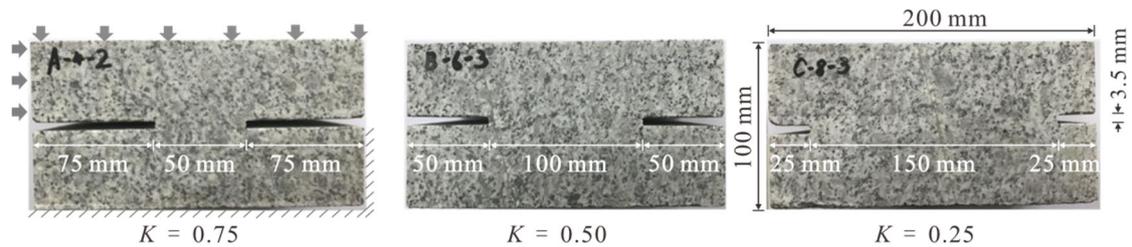

**Fig. 3** Granite specimens with different joint persistence $K$ = 0.75, 0.50 and 0.25

UDEC was adopted to simulate the above experiments. The numerical model used to calibrate the uniaxial compression tests had a width of 50 mm and height of 100 mm and was generated by UDEC trigon approach to 4,409 blocks bonded by 25,522 contacts with an average 1.5-mm block edge length, which had been proven to be low enough to have less influence on the simulated failure pattern. Upper velocity boundary of 0.01 m/s was applied considering the amount of calculation (Gao and Stead 2014).

To calibrate the direct shear tests of rock bridges, another numerical model with a width of 200 mm and a height of 100 mm was generated using the modified UDEC trigon approach. Inspired by Tatone and Grasselli (2015) and Kemeny et al. (2016), the block size linearly increased as a function of the distance from the shear interface to underline the region of rock bridges. Furthermore, this numerical model was cut into three different models with the joint persistence of 0.75, 0.50 and 0.25, with correspondence to 8,876 blocks with 57,163 contacts, 9,030 blocks with 58,512 contacts and 9,204 blocks with 59,906 contacts. According to the physical prototypes, the lower parts of the numerical models were fixed, and a constant normal stress boundary of 4 MPa was applied until the numerical models were stable (the maximum unbalanced force was less than $10^{-3}$ N). Then, this normal stress boundary was kept invariant, and a left velocity boundary of 0.01 m/s was added to the upper parts.

In terms of calibration, multiparameter sensitivity analysis was introduced to set the default initial microparameters and their possible ranges. Next, based on the physical prototype constraints, the microparameters and ranges in UDEC would be further determined. Finally, these parameters would be calibrated according to previous macro-microscopic parameter relationships until the numerical results were generally in line with the physical prototype results. In this study, the microparameters shown in Table 1 had been calibrated for a range of macroparameters of both uniaxial compression tests and direct shear tests presented in Table 2, indicating that the simulation in UDEC agreed well with the prototype results.

| Properties | Values |
|---|---|
| Young's modulus of blocks, $E^{block}$ (GPa) | 27.20 |
| Poisson's ratio of blocks, $\mu^{block}$ | 0.12 |
| Normal stiffness of contacts, $k_n$ (GPa/m) | 140,600 |
| Shear stiffness of contacts, $k_s$ (GPa/m) | 56,240 |
| Contact cohesion, $c^{cont}$ (MPa) | 24.80 |
| Contact friction angle, $\varphi^{cont}$ (°) | 23.00 |
| Contact tensile strength, $\sigma_t^{cont}$ (MPa) | 4.00 |

**Table 1** Calibrated micro-properties used in UDEC to present the granite specimens

| Properties | Experiment | Modelling |
|---|---|---|
| Young's modulus, $E$ (GPa) | 21.48 | 21.53 |
| Poisson's ratio, $\mu$ | 0.16 | 0.16 |
| UCS (MPa) | 87.80 | 87.75 |
| Cohesion strength, $c$ (MPa) | 23.40 | 21.39 |
| Internal friction angle, $\varphi$ (°) | 33.50 | 32.08 |
| Shear strength $K_{0.75}$ (MPa) | 10.74 | 10.63 |
| Shear strength $K_{0.50}$ (MPa) | 14.16 | 13.82 |
| Shear strength $K_{0.25}$ (MPa) | 15.53 | 16.16 |

**Table 2** Calibrated results of macro-properties

## 2.3 Numerical models and monitorings

As shown in Fig. 4, the three above numerical models used for calibration were extended to 18 models with different numbers of rock bridges, $n$, while maintaining a constant joint persistence of 0.75, 0.50 and 0.25. The $n$-value was set as integers from 1 to 6, meaning that the intact rock bridge was dispersed into equivalent lengths. To maintain the comparability, the dispersed part should keep the same total length of joints and rock bridges, and boundary conditions were the same as those in the calibration process. In this way, the dispersion of rock bridges became the only independent variable.

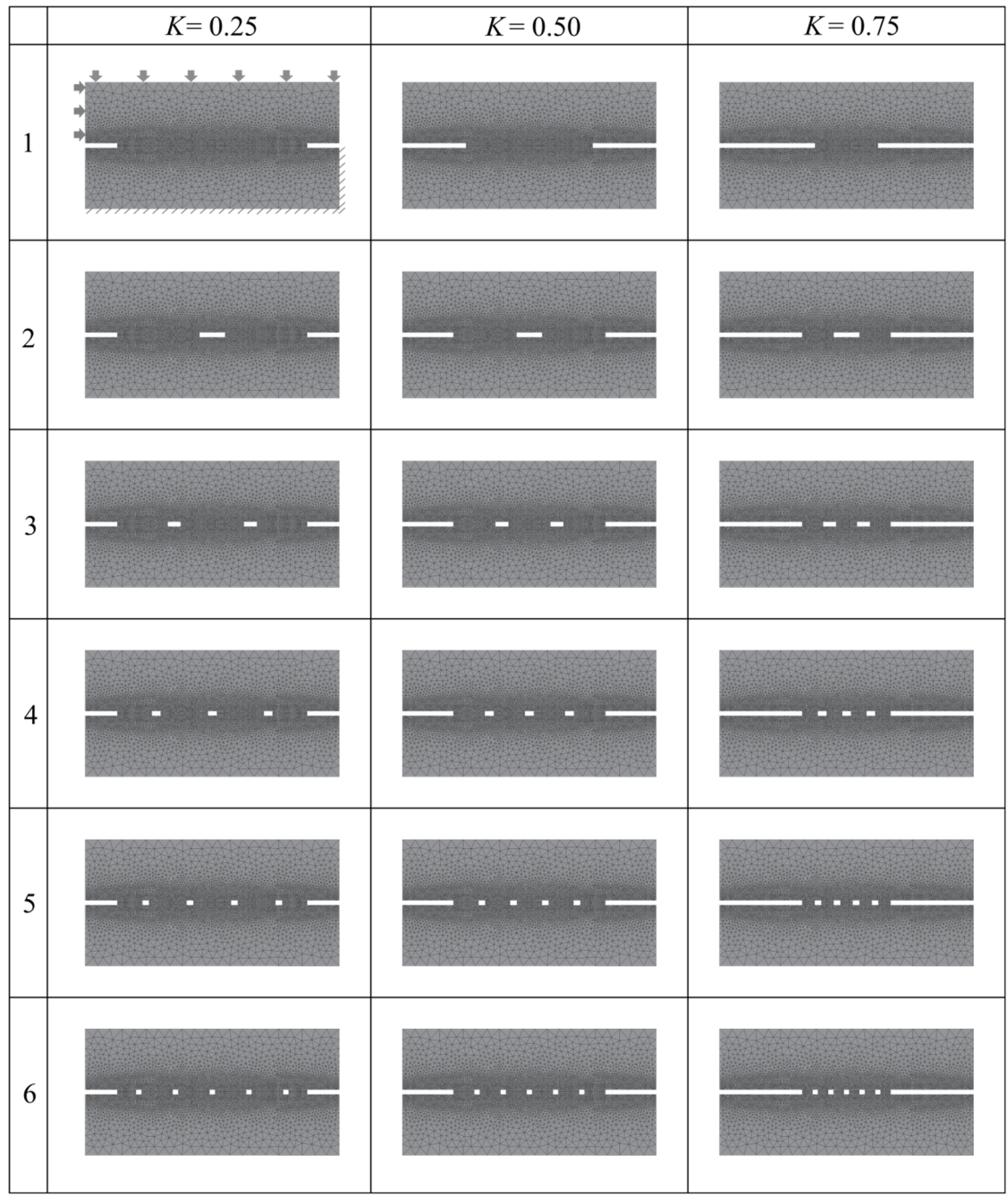

**Fig. 4** Numerical models with rock bridges from 1 to 6 in number while maintaining constant joint persistences $K$ = 0.75, 0.50 and 0.25

In the simulation process, the monitorings of shear load (reaction force on the horizontal velocity boundary), stress and displacement fields were implemented by writing various FISH functions. In addition, it is necessary to introduce the simulated acoustic emission (AE) to characterize the failure processes of rock bridges. Typically, Bu et al. (2021) proposed a feasible and simple model to simulate AE by monitoring and analyzing element velocity in DEM, which was applied in this research to acquire AE events and AE energy.

## 3 Results

### 3.1 Peak shear resistance

Figure 5 shows the variation in shear load versus shear displacement (horizontal displacement parallel to the shear load) with different joint persistence ($K$ = 0.25, 0.50 and 0.75) and different dispersion ($n$ = [1,6]). Acquired 18 curves showed similar variation trends, that is, with increasing displacement, the shear load increased in an approximately linear manner at first, then, the increasing rate of load decreased gradually before reaching the peak shear resistance, after that, the curves lied in postpeak stages, and the load dropped significantly. While differences among curves with different joint persistence were conspicuous, as expected, the peak shear resistance decreased with increasing joint persistence because of the increasing stress concentration (Ghazvinian et al. 2012). These differences would not be discussed further in this paper, and more attention was paid to the differences among scenarios with various dispersions with the same joint persistence.

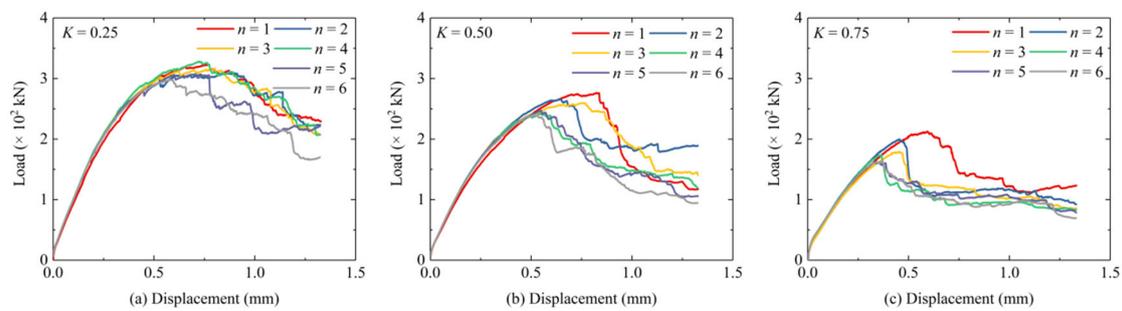

**Fig. 5** Load versus displacement of numerical models with different dispersion ($n$ = [1,6]) of rock bridges and different joint persistence **a** $K$ = 0.25; **b** $K$ = 0.50; and **c** $K$ = 0.75

As shown in Fig. 6, the peak shear resistance decreased with increasing dispersion in an approximately linear manner, and the absolute decreasing slope was directly proportional to the joint persistence. Precisely, the absolute slope of 0.35 kN, 0.73 kN and 1.03 kN corresponded to the joint persistence of 0.25, 0.50 and 0.75, respectively. In addition, as shown in Fig. 7, the peak shear displacement also decreased with increasing dispersion. The above results were consistent with the experimental studies of scale effects on the shear behaviours of rock joints (Bandis et al. 1981).

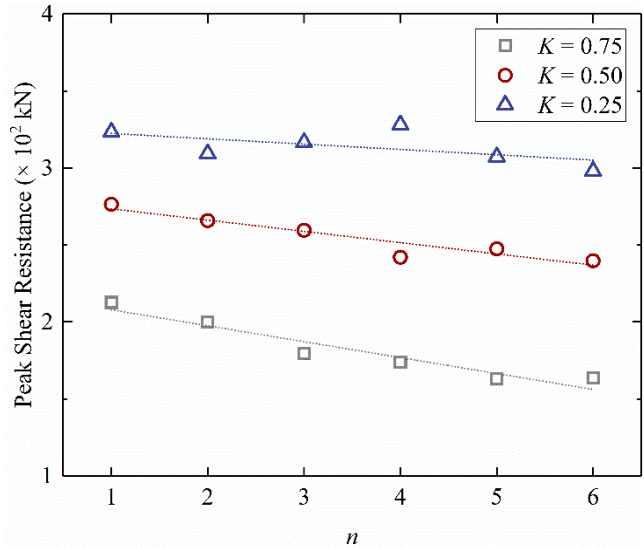

**Fig. 6** Peak shear resistance versus dispersion ($K$ = 0.50; $n$ = [1,6])

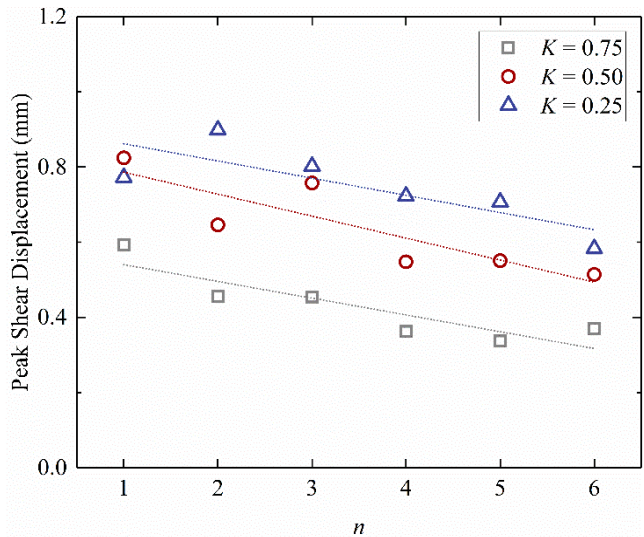

**Fig. 7** Peak shear displacement versus dispersion ($K$ = 0.50; $n$ = [1,6])

### 3.2 Shear stress field

Progressive failure processes of rock bridges have been widely claimed to be due to the stress concentrations at the tips of pre-existing cracks (Lajtai 1969; Jiang et al. 2006; Yang et al. 2009; Zhang and Wong 2012; Fan et al. 2018), thus, it is necessary to investigate the stress field of rock bridges. Numerical simulations with a joint persistence of 0.50 were taken as an example to interpret the scale effects. When the numerical models were subjected to a constant confined compression without shear load and were stable, as shown in Fig. 8, the scale effects of rock bridges on the normal stress ($\sigma_y$) field were appreciable. Although significant stress concentrations at the tips of the end pre-existing joints were similar in number and shape, the stress concentrations of the inside rock bridges among six numerical models showed differences that the concentrated region and value decreased with increasing dispersion, indicating that the end rock bridges bore the key confined compression, while the inside ones shared it almost equally.

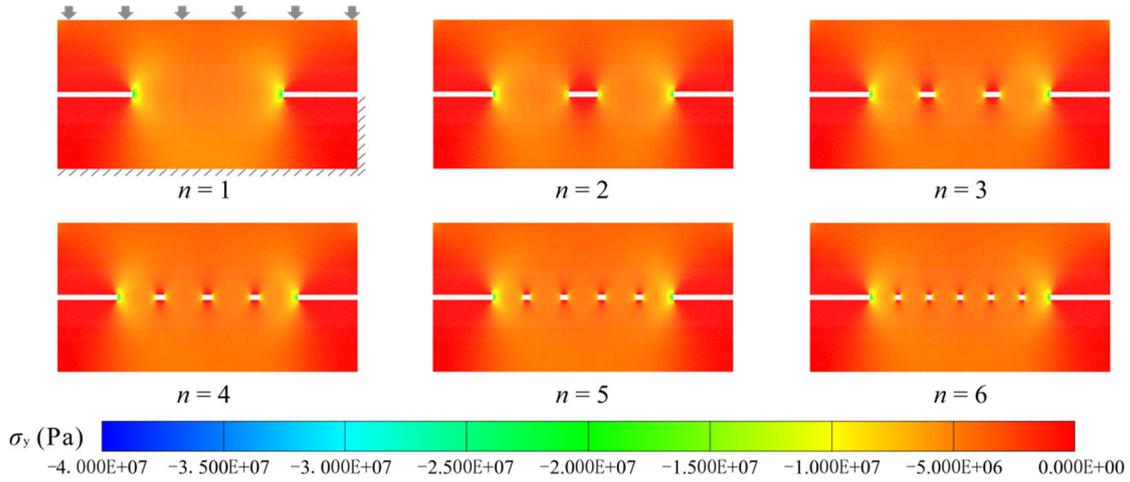

**Fig. 8** Normal stress ($\sigma_y$) field distribution of numerical models with different dispersion while keeping the same joint persistence of 0.50 when the numerical models were stable in compression without external shear load

Upon shearing, the orientation of the principal stress ($\sigma_1$) would experience a transition from normal ($\sigma_y$) to lateral ($\sigma_x$). Fig. 9 showed the shear stress ($\tau_{xy}$) field states of prepeak stages (displacement of 0.20 mm), approximate peak stages (displacement of 0.80 mm) and postpeak stages (displacement of 1.33 mm). During the prepeak stages, $\tau_{xy}$ distribution of rock bridges was not uniform, and significant stress concentration appeared at the upper tips of left joints and the lower tips of the right joints, which was highly consistent with analytical results by Segall and Pollard (1980) and experimental results by Allersma (2005). In terms of multi-rock bridges, these above phenomena remained visible, while the concentrated region and value decreased with increasing dispersion.

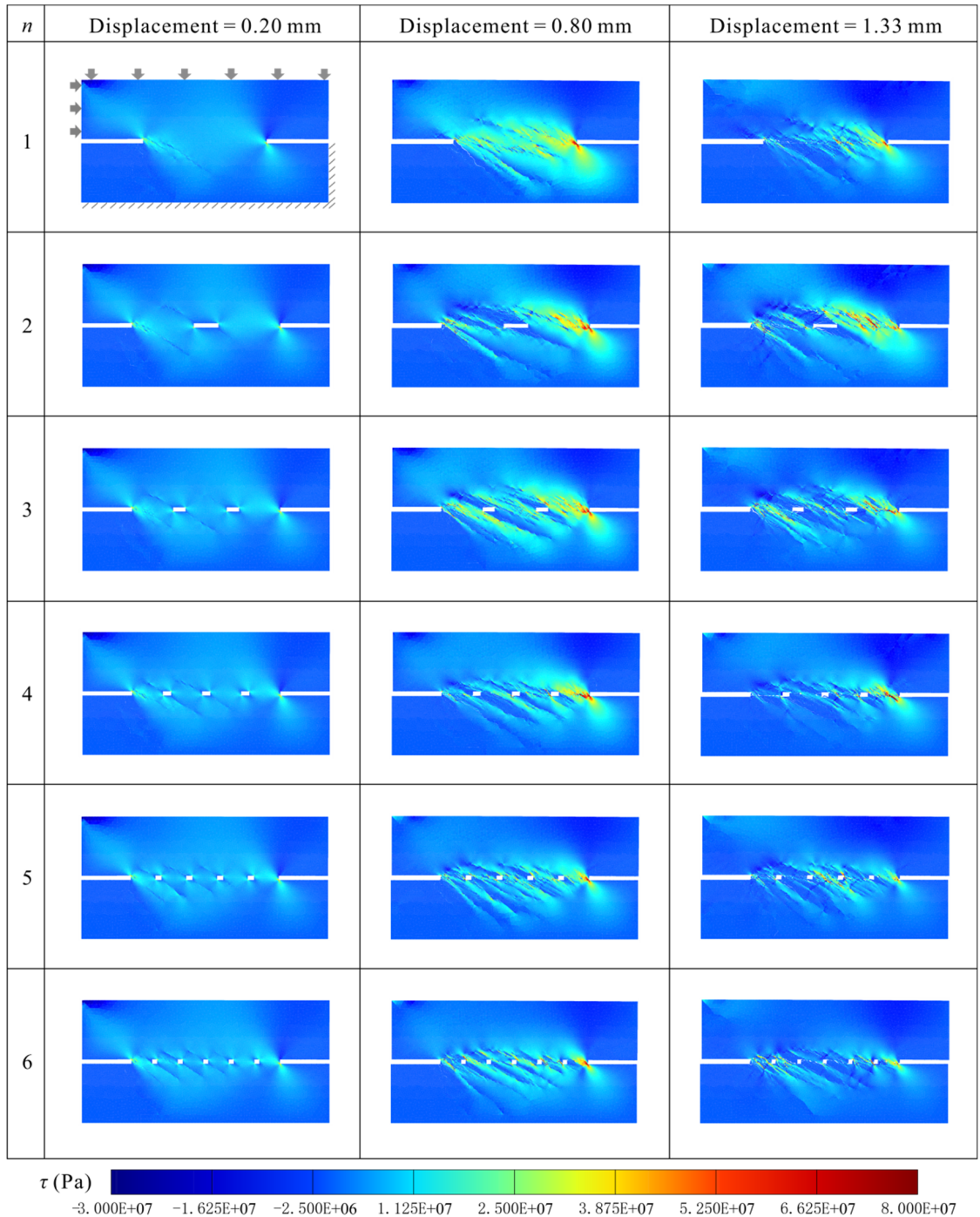

**Fig. 9** Shear stress ($\tau_{xy}$) field distribution of numerical models with different dispersion while keeping the same joint persistence of 0.50 at the displacement of 0.20 mm, 0.80 mm and 1.33 mm

After prepeak stages, the load-displacement curves of the numerical models reached the peak-stress points. As shown in Fig. 9, significant stress concentration appeared at the tips of the rock bridges further from the shear load, while the area of the concentrated region decreased with increasing dispersion. By contrast, rock bridges closer to the shear load bore less shear resistance compared with that at prepeak stages because most of these rock bridges had been broken, concluded from the section "Displacement field".

When the numerical models lied in postpeak stages, the bearing capacity dropped significantly, manifested by smaller stress concentration values and areas. The concentrated positions were also at the tips of the rock bridges further from the shear load, and the concentrated values and areas decreased with increasing dispersion as well.

### 3.3 Displacement field

Compared with stress field monitoring, displacement field is also important but rare in rock bridge domain. It matters because the displacement field can intuitively reflect the transfer of inside strain, reasonably interpret the macro failure modes (Sarfarazi et al. 2014; Zhang and Wong 2014) and better explain the crack initiation and propagation (Cao et al. 2017). As shown in Fig. 10, the displacement magnitude fields variated with the states of numerical models, and this uneven distribution and evolution of the displacement fields were due to the presence of pre-existing rock bridges according to Paronuzzi et al. (2016).

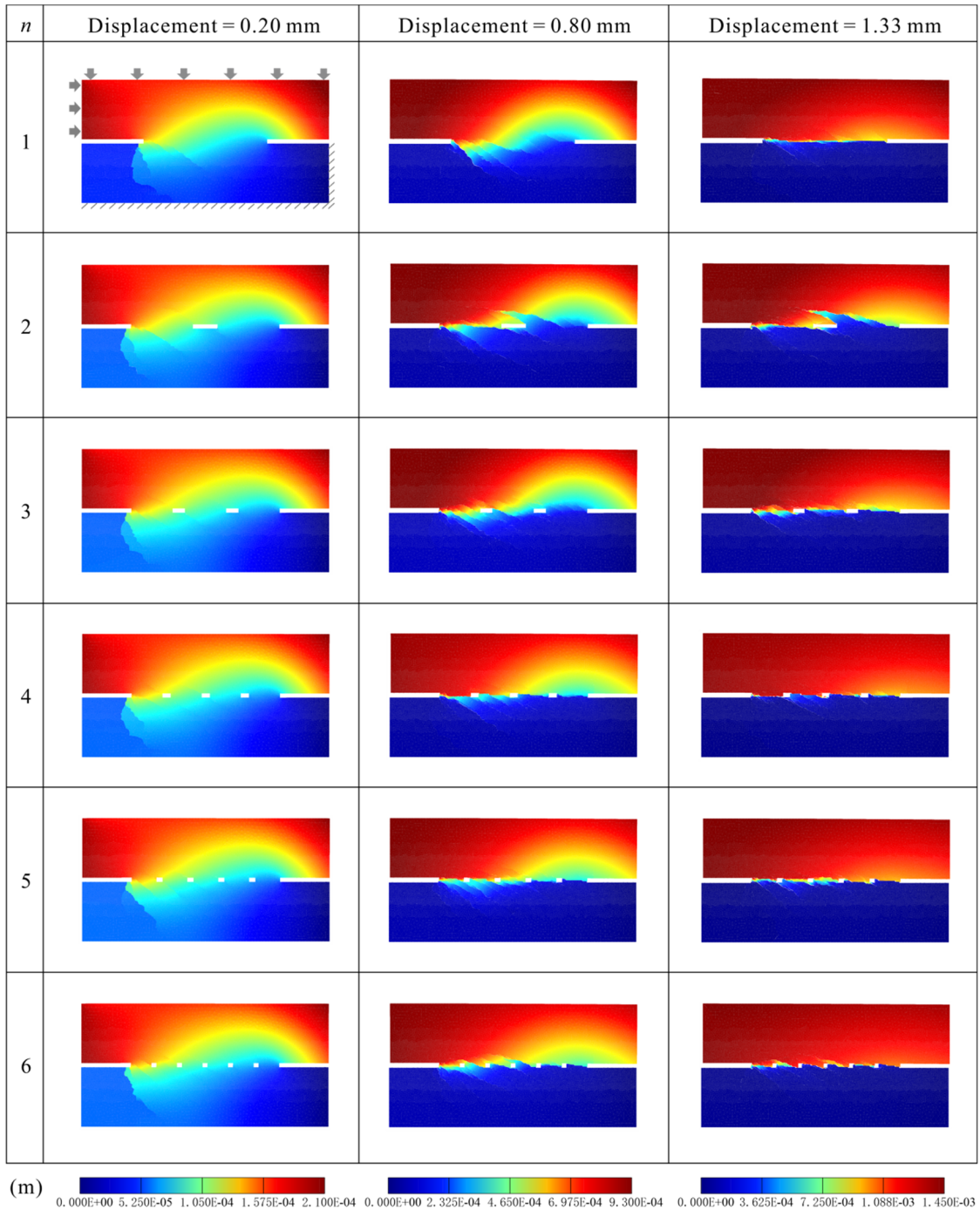

**Fig. 10** Displacement magnitude field distribution of numerical models with different dispersion while keeping the same joint persistence of 0.50 at the displacement of 0.20 mm, 0.80 mm and 1.33 mm

At the prepeak stages, an intuitive result was the arc distribution of displacement, consistent with the research by Cundall et al. (2016). This arc formed at the position of rock bridges and had transmitted away from the load. The shapes and areas of arc among models of different dispersion were similar.

Other characteristics of this arc were the asymmetry and approximate continuity, of which the former was due to the progressive failure, and the latter was greatly affected by the pre-existing joints and formed microcracks. These formed microcracks could be clearly observed because the displacement field would vary abruptly in magnitude within a short distance in the vicinity of a crack (Zhang and Wong 2014). As shown in Fig. 10, clear tensile microcracks that initiated from the tips of the left joints were observed, and the propagation was constrained by Griffith theory, that is, tensile wing cracks propagated in the direction normal to the axis of pre-existing joints firstly, followed by the extension along a curvilinear path that aligned with the direction of the principal stress, which had been proven to be a typical crack pattern in rock bridges by Sarfarazi and Haeri (2016). This propagation pattern of tensile wing cracks was hardly affected by scales of rock bridges. In addition to the wing cracks, another typical mode was secondary cracks described as shear zones formed by the accumulation of microscopic tensile cracks that initiated in a direction coplanar or quasi-coplanar to the pre-existing joints (Ghazvinian et al. 2012). While the propagation of secondary cracks at this stage had some differences that the length decreased with increasing dispersion.

When the load-displacement curves of the numerical models approximately reached the peak-stress points, the position, shape and area of the arc and crack propagation changed appreciably. In terms of the arc, it moved away from the load further, and the displacement field on the left side of the arc varied significantly, indicating the failure of left rock bridges that were closer to the load. Precisely, half of the rock bridge was broken significantly, which was common in all of the six models. Another change was that the area of the arc became smaller, while the reduction increased with increasing dispersion. In terms of the crack propagation, tensile wing cracks at the tips of left rock bridges were not clear because the range of displacement contours increased several times, in other words, displacement in the shear zone was greatly larger than the displacement at tensile wing cracks. Thus, secondary cracks were visually main pattern at this stage. Furthermore, left formed fracture had different morphology that the asperity basically decreased with increasing dispersion, since the smaller intact rock bridges would be easier to be broken in shear.

Displacement fields at postpeak stages were also shown in Fig. 10. The arc at the previous two stages was invisible, and the displacement fields strongly separated along the rock bridges, indicating that the cracks had a large-scale coalesce. Formed rough macro fracture provided residual shear strength, and the roughness basically decreased with increasing dispersion as well.

### 3.4 AE characterization

AE characterization is becoming a popular interest in rock bridge study (Chen et al. 2015; Guo et al. 2020; Jiang et al. 2020; Qin et al. 2020; Wu et al. 2020; Zhang et al. 2020). Normally, AE signals accompanied with fracturing events were always operated in the form of waveform analysis to acquire AE parameters to quantify the failure process of rocks, of which the AE events and AE energy were classical AE parameters. To better interpret the progressive failure processes of rock bridges, the average $\tau_{xy}$ of each rock bridge was also recorded. As shown in Fig. 11a, there was a sharp increase of both AE events and AE energy before and near the peak strength point, and these two points could be respectively determined as the volume-expansion point and the peak-stress point according to Moradian et al. (2016), which had been proven to play a crucial role in the evolution of damage in heterogeneous rocks and to predict the macroscopic rupture (Xue et al. 2014a; Xue et al. 2014b; Chen et al. 2021). At the volume-expansion point, both curves of the shear load and $\tau_{xy}$ of rock bridges emerged with slight drops, meaning that the model was perceptibly fractured. Moving to the peak-stress point, both curves dropped significantly just before the second intensive AE activities, which was claimed to be the lag of macro fracture behind the peak-stress point (Zhu et al. 2019).

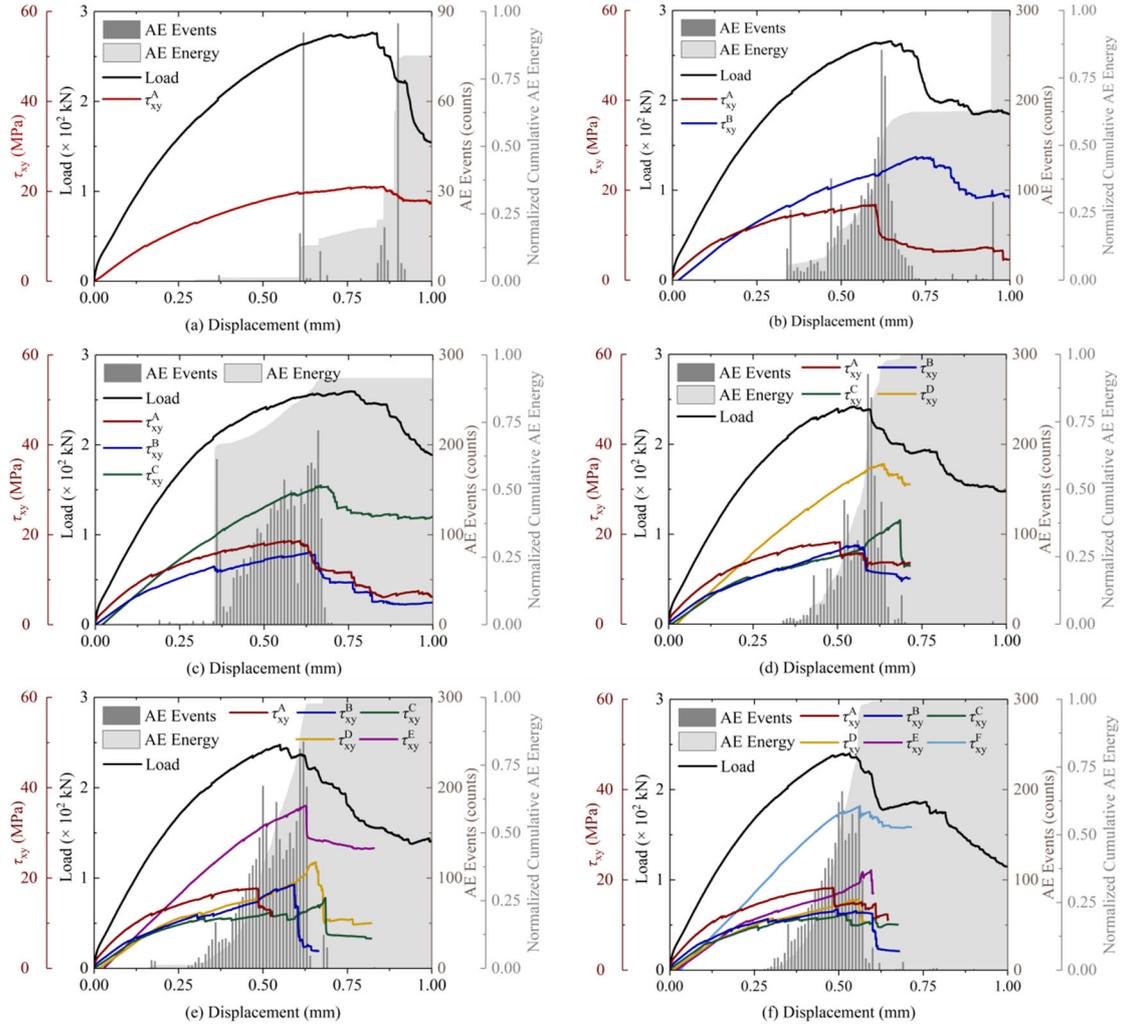

**Fig. 11** Average shear stress ($\tau_{xy}^\eta$), load and AE parameters (AE events and energy) versus displacement with different dispersion ($n = [1,6]$) while keeping the same joint persistence of 0.50, where η was the left-to-right serial number

In addition to the intact rock bridge, multi rock bridges had similar characteristics in curves of $\tau_{xy}$. Although average $\tau_{xy}$ might not be an extremely accurate index to present the damage of each rock bridge, it could reflect the failure sequence and the transmission of stored strain energy to a certain degree. As shown in Fig. 11a~f, all of the $\tau_{xy}^A$ curves were the first to occur significant drops, and the last plunge always happened at the $\tau_{xy}$ curves of the final (rightmost) rock bridge, agreeing with the deduction using stress and displacement fields. Furthermore, the last rock bridges commonly had the strongest resistance, almost two times larger than other rock bridges. These phenomena illustrated that the left rock bridge closer to the load end was prior to fracture, releasing partial strain energy meanwhile. And another part of the strain energy was transmitted and stored to the right unbroken rock bridges further from the load end. With more stored energy, the last rock bridges provided key resistance, meaning that the breakage of the last rock bridges would result in the macro failure of the whole model.

The intensive AE activities always indicated fracturing events on a relatively large scale. Several sudden increases in AE events and AE energy could be observed in Fig. 11, of which the last increase was the highest, corresponding to the peak-stress point and significant drops of $\tau_{xy}$ curves of the final (rightmost) rock bridges. It was also valuable to analyse the intensive AE activities at prepeak stages as the precursors to macroscopic rupture. As shown in Fig. 11, the precursor of intact rock bridge was easier to be identified compared with the precursors of multi rock bridges. With increasing dispersion, the shear resistant zone was more crushed, leading to a superposition of the precursors.

## 4 Discussion

### 4.1 Concept of the rock bridge: Insights from scale effects

Elmo et al. (2018) proposed several important but unanswered questions related to rock bridges, and the first is "What is a rock bridge and what parameters govern whether a given intact rock portion of the rock mass can or should be defined as a rock bridge?" To answer this question, considering that the shear resistance of rock bridge would be reduced with higher fragmentation until the original intact rock bridge was deprived of the resistance and could not be defined as a rock bridge, we designed an intact rock bridge in shear and dispersed it into different scales but kept the joint persistence invariant. In this way, there should be a scale threshold between rock bridge and non-rock bridge. According to section "Peak shear resistance", there was a linear relationship between the peak shear resistance and dispersion. Nevertheless, when the dispersion was extended further, as shown in Fig. 12, the absolute slope decreased gradually. In particular, a relatively significant transition was observed when $n$ fell in the range of [9,10], after which the absolute slope perceptibly decreased. In this case, the joint persistence of a unit (defined as a combination of an individual rock bridge and an adjacent joint) was in the vicinity of 0.20. Interestingly, according to the report by Diederichs and Kaiser (1999), this was a threshold that cracks began to act significantly in mechanical properties, which might be a possible value to define a rock bridge. Besides, as shown in Fig. 12, perceptible fluctuations could be observed when $n$ was more than 12 in number since with smaller scales, grain size became important enough to affect the rock bridge responses (Schultz 1996), and more dispersion would be questionable.

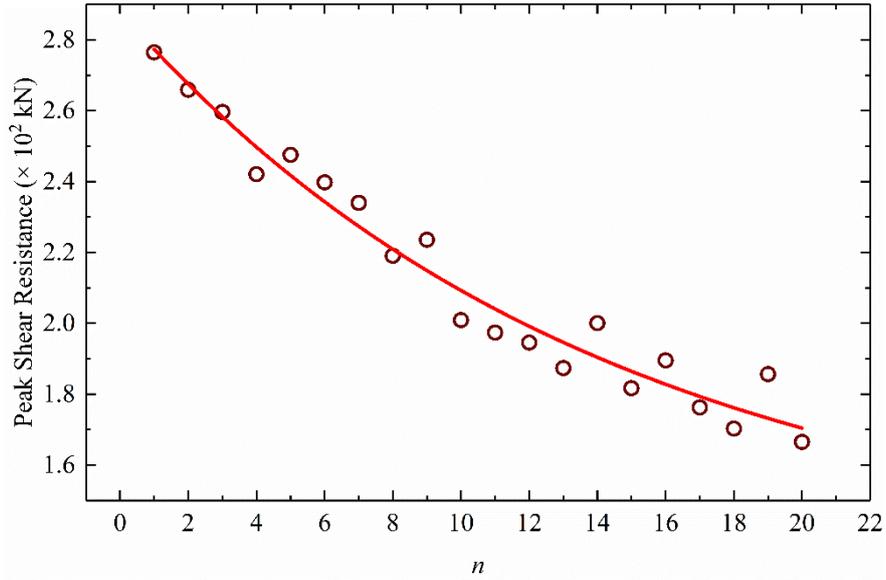

**Fig. 12** Peak shear resistance versus dispersion ($K = 0.50$; $n = [1,20]$)

**4.2 Concept of the joint persistence: Insights from scale effects**

As noted by this research and previous understandings (Shang et al. 2018; Tang et al. 2021), scale effects of rock bridges presented a challenge to the concept of the joint persistence. Based on a traditional back analysis method, we assumed that an intact rock bridge was dispersed as multi rock bridges while maintaining a constant joint persistence, which was in an ideal state. Thus, it seemed to be unreasonable to propose a new criterion according to this assumption to replace the concept of the joint persistence. With regard to the rock engineering design, joint persistence had been widely concerned in the rock mass classification systems. Excellent rock mass classification systems were RMR (Bieniawski 1973), Q (Barton et al. 1974) and GSI (Hoek and Kaiser 1995). These classical and inductive approaches were always based on the experience and subjectivity of engineers, leading to potential inevitable errors in rock mass characterization (Elmo et al. 2021a). According to the report by Kim et al. (2007), joint persistence was involved in RMR system, but its weighted values were underestimated; Q system gave insufficient consideration to the joint persistence in practice; and GSI descriptively referred joint persistence as block interlocking. In this context, several researchers made positive contributions to refer joint persistence in rock mass classification systems. Laubscher (1990) proposed a measurement using the average fracture frequency rating (FFR) in MRMR classification system. The rock bridge percentage $P_{rb}$ was determined as follows:

$$P_{rb} = \frac{FFR \times a}{40} \tag{7}$$

where *a* was the constant and could be determined using experiences (Dempers et al. 2011). While this measurement was pointed to neglect the fundamental role of fracture persistence (Elmo et al. 2018). Furthermore, Elmo et al. (2021a) proposed a network connectivity index (NCI) based on a combination of fracture intensity, fracture density and fracture intersection density parameters, expressed by the formula:

$$\text{NCI} = \frac{P_{21}}{P_{20}} I_{20} \tag{8}$$

where $P_{21}$, $P_{20}$ and $I_{20}$ were the areal fracture intensity, the areal fracture density and areal fracture intersection density, respectively. Then, Elmo et al. (2021b) introduced the concept of $\text{NCI}_d$ by considering the impact of stress-driven fractures, defined by:

$$\text{NCI}_d = \text{NCI} + \text{NCI}_{rb} \tag{9}$$

where $\text{NCI}_{rb}$ was the NCI calculated considering just the induced fractures. The ratio between $\text{NCI}_{rb}$ and $\text{NCI}_d$ was defined as the rock bridge potential (RBP), as shown in Fig. 13 and Eq. 10:

$$\text{RBP} = \frac{\text{NCI}_{rb}}{\text{NCI}_d} = \frac{\text{NCI}_{rb}}{\text{NCI} + \text{NCI}_{rb}} \tag{10}$$

In this way, the higher RBP meant the higher contribution of rock bridge to the overall rock mass behaivours. In this study, we adopted RBP index to represent the scale effects of rock bridges, as shown in Fig. 14.

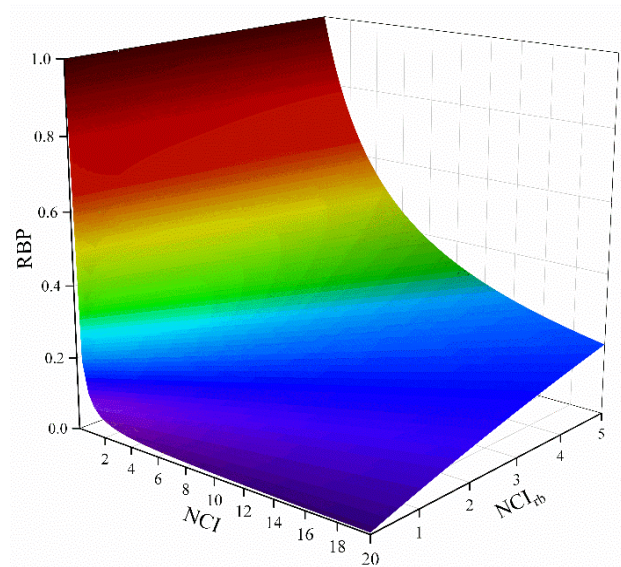

**Fig. 13** Definition of the RBP as a function of the initial NCI and the induced $\text{NCI}_{rb}$

In addition, some other calculations of joint persistence were proposed as well. For example, Park (2005) proposed a probabilistic analysis by treating joint persistence as a random variable, implemented by comparing the individual joint length with the maximum sliding dimension in each interval on the premise that the discontinuity on the failure plane was fully persistent. Wasantha et al. (2014) also proposed an index named the degree of persistence (DoP) by considering the impact of discontinuous joint tips on stress distribution characteristics:

$$\text{DoP} = \frac{\sum J_i}{n_d \times (\sum J_i + \sum R_i)} \quad (11)$$

where $J_i$ and $R_i$ were the ith length of the joint and the rock bridge, and $n_d$ was the number of discontinuous joint tips. DoP index was also used to characterize the scale effects of rock bridges and was compared with RBP, as shown in Fig. 14. Although the span was different, both indexes showed consistent feasibility in describing the scale effects.

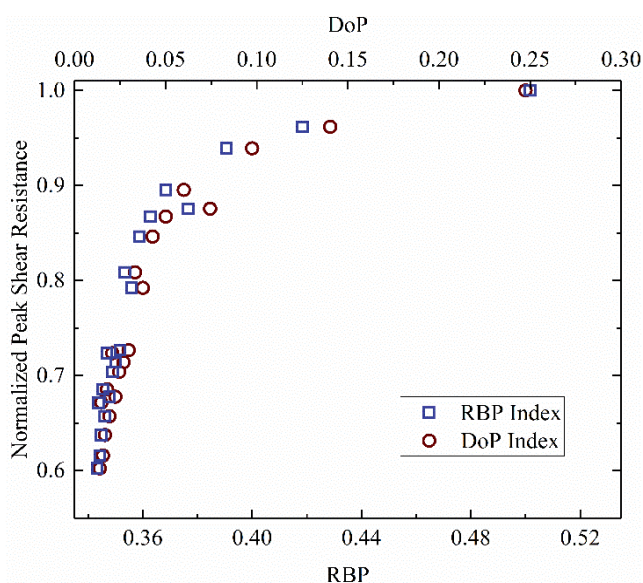

**Fig. 14** Normalized peak shear resistance versus RBP index and DoP index

**4.3 New insights into the scale effects of rock bridges**

Our research revealed erroneous equivalent rock mass responses in detail when using the concept of joint persistence. With invariant joint persistence, mechanical properties of rock bridges deteriorated with increasing dispersion, manifested by lower shear resistance, lower stress-concentration area and value, higher displacement arc field reduction and lower asperity of macro shear fracture. A negative exponent relationship between the shear resistance and dispersion was found, in which the joint began to act significantly to the shear resistance at the joint persistence of 0.20, and this was possible to be used to define a rock bridge. Besides, both RBP and DoP were appropriate indexes when considering the scale effects of rock bridges.

### 4.4 Limitations of the present study

In the present study, numerical simulations were implemented at the 2D level, bringing the limitations of high consistency with the 3D physical prototypes though the results at the 2D level might theoretically appear compatible at the 3D level and could save plenty of time to calculate and analyse. Furthermore, this research focused on filling in the gap of scale effects of rock bridges, and other factors, such as minerals, joint length and orientation, normal stress and loading rate, were not included. In addition, our understandings of the scale effects of rock bridges were still at the laboratory scale and based on a traditional back analysis method due to the invisibility of rock bridges, which might be improved by introducing some techniques to identify rock bridges in the field.

### 5 Conclusions

To fill in the gap of the investigation of the scale effects of rock bridges, numerical simulations were performed on direct shear tests with different dispersion of rock bridges using UDEC while maintaining a constant joint persistence. From the perspective of load-displacement curves, stress and displacement fields, crack propagations and AE characterizations, the following key conclusions could be drawn.

1. Both shear resistance and peak shear displacement decreased with increasing dispersion, and the former reduction increased with increasing joint persistence. While the shear modulus was insensitive to the scale of rock bridges.
2. Rock bridges at the end bore the key resistance, and stress-concentration area and value decreased with increasing dispersion.
3. The uneven distribution of displacement field of rock bridges was in an arc manner moving and degrading away from the load, illustrating the sequent failure of multi rock bridges. The relative degradation of this arc increased with increasing dispersion.
4. Propagation of wing cracks was not sensitive to the scale, while the asperity of macro shear fracture mainly formed by secondary cracks basically decreased with increasing dispersion.
5. Dispersion of rock bridges might lead to the overlap of the precursors identified by intense AE events and abrupt AE energy.
6. A unit with joint persistence of 0.20 might be a threshold to define a rock bridge, and both RBP and DoP were appropriate indexes when considering the scale effects of rock bridges.